# Human-Centric Data Cleaning [Vision]

El Kindi Rezig*    Mourad Ouzzani°    Ahmed K. Elmagarmid°    Walid G. Aref*

*Purdue University    °Qatar Computing Research Institute
erezig@cs.purdue.edu, mouzzani@hbku.edu.qa, aelmagarmid@hbku.edu.qa,
aref@cs.purdue.edu

## ABSTRACT

Data Cleaning refers to the process of detecting and fixing errors in the data. Human involvement is instrumental at several stages of this process, e.g., to identify and repair errors, to validate computed repairs, etc. There is currently a plethora of data cleaning algorithms addressing a wide range of data errors (e.g., detecting duplicates, violations of integrity constraints, missing values, etc.). Many of these algorithms involve a human in the loop, however, this latter is usually coupled to the underlying cleaning algorithms. There is currently no end-to-end data cleaning framework that systematically involves humans in the cleaning pipeline regardless of the underlying cleaning algorithms. In this paper, we highlight key challenges that need to be addressed to realize such a framework. We present a design vision and discuss scenarios that motivate the need for such a framework to judiciously assist humans in the cleaning process. Finally, we present directions to implement such a framework.

## 1. INTRODUCTION

Businesses often collect large volumes of data to inform key decisions. However, because data can be humongous and highly volatile, it is infeasible for humans to manually verify its accuracy. As a result, decision-makers have to deal with possibly-inaccurate data that may inherently lead to faulty business decisions.

There are abundant research efforts to detect and repair the many types of data errors that one sees in the wild. This process is also known as the data cleaning process. Data errors include duplicates [9], violations of integrity constraints [5], and missing values [2]. While ideally we want to be able to fully automate this process, it has been widely recognized that humans have to be involved at various stages of the data cleaning process. A large spectrum of data cleaning systems involve humans. Examples include Poter's Wheel [16], GDR [22], KATARA [6], CrowdER [20], and UGuide [18]. Each of these systems involves humans to solve a particular data cleaning task. However, an end-to-end data cleaning framework that involves humans in a way that is orthogonal to the underlying cleaning algorithms is not yet available. Looking at existing data cleaning techniques, we make the following observations:

**Human involvement is algorithm-driven:** Humans are the ultimate authority in verifying the accuracy of the data. Because it is impractical to have humans correct the entirety of the data, many techniques strive to involve humans judiciously so as to maximize the benefit of their feedback in the cleaning process [22, 20, 21, 15]. Typically, humans are tightly coupled to the cleaning logic, i.e., humans are involved in ways that are dictated by the cleaning algorithm being used. This coupling precludes a generic inclusion of humans in the cleaning pipeline regardless of the underlying cleaning algorithms being used.

**Data singularity:** An assumption that is usually made in cleaning algorithms is that the data is subjected in its entirety to a given cleaning algorithm [8, 17]. However, in practice, different parts of the data are cleaned by different agents. For instance, one may use an automatic data cleaning tool to find the correct mapping of the Zip code to a City name while requesting humans to correct the Salary data (one would not trust an automatic algorithm to modify the salary data). This motivates the need for a cleaning framework that supports both human and automatic agents to holistically clean different parts of the data. In the remainder of the paper, we refer to the detection and repairing agents as *cleaning agents*.

**Source of errors:** There are multiple factors that can influence the quality of the computed repairs, e.g., the humans involved, the data quality rules, and the repair algorithm. However, existing techniques do not assess the effect of various factors that produce the data repairs (for example, typical rule-based repair algorithms assume the rules are correct [7, 4, 13]). Understanding this effect is crucial in identifying bottlenecks in the data cleaning pipeline. Just as important as suggesting potential data errors for humans to verify, it is important to make it easy for humans to identify *faulty* factors (rules, humans, external resources, etc.) that have been involved to compute the inaccurate repairs.

**Humans are not always right:** Many human-driven data cleaning techniques assume that humans (e.g., experts) are *perfect* [22]. However, in practice, humans may make mistakes at various stages of the data cleaning pipeline (e.g., in the detection, repairing, or validation phases). Understanding how humans interact with the data is important to judiciously involve them in the cleaning process. For in-stance, an error reported in the *Sales* data by a person working in the *Sales* department should have more weight than one reported by a human working in another department. Therefore, there are several nuances in the human feedback that need to be dissected to effectively involve humans in the cleaning process.

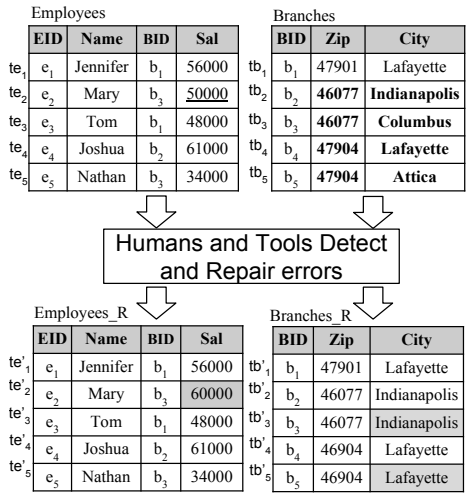

Figure 1: Example tables

*Example 1.* Refer to Figure 1. Table *Employees* contains employee data, e.g., name, salary, and the branch they belong to (BID). Table *Branches* contains the list of branches (BID) and their location (Zip, City). We assume all branches are based in the State of Indiana, thus we omit the State attribute from Table *Branches*.

**Scenario 1:** Assume that a human and an automatic agent participate to detect and repair the data. The question is: How to effectively involve humans in the cleaning process?

**Detection:** Assume that a human, say *Alice*, reports a data error in the salary of employee $e_2$ (the salary value is underlined in Figure 1). Additionally, consider an FD rule $\phi_1$: $Zip \to City$ that is defined on Table *Branches*. $\phi_1$ states that records sharing the same Zip code must share the same City name. The data values marked in boldface in Figure 1 correspond to violations of $\phi_1$.

**Repairing:** Assume that there is another human, say *Bob*, who is assigned to correct the error reported by *Alice*. The violation of $\phi_1$ is repaired using an automatic FD repair tool, say $R_1$. In Figure 1, Tables *Employees_R* and *Branches_R* contain the computed repairs (the shaded cells correspond to cells that have been updated by the repair agents). Scenario 1 raises several questions:

- How do we know how confident *Alice* is about the reported Salary value? Asking another human, in this case *Bob*, to verify the reported errors is expensive. How can we model *Alice's* knowledge on different parts of the data?

- How do we assess *Bob's* ability to fix the Salary value reported by *Alice*? In other words, if another human, say *Sam*, who claims he could also fix the Salary value, how does the system choose between the fixes of *Bob* and *Sam* to the value of Salary? The system can also ask both of them. On one hand, asking either one of them is efficient but should only be performed if the chosen person has enough confidence about fixing the Salary value. On the other hand, asking both of them is expensive but can be necessary if they are both known to make a few mistakes when fixing Salary values.

- What if $R_1$ produces inaccurate repairs. Should we blame $R_1$ or $\phi_1$? Notice that both of these parameters come into play to produce the FD repairs. What if the corresponding FD rules that support the repairs are incorrect or that the repair algorithm produces probabilistic or heuristic output? How can the framework isolate the *culprits* in the data cleaning pipeline?

The above questions are among many others that need to be addressed to effectively involve humans in the data cleaning pipeline. To this end, we propose a vision for an end-to-end data cleaning system that supports the following features:

- **Heterogeneity:** The system should be able to simultaneously support cleaning agents of different types, i.e., human, automatic or semi-automatic. Since different parts of the data can be cleaned by different agents, each agent receives part or all of the data as input.

- **Isolation:** The system should treat cleaning agents as black boxes while still enabling humans to detect, repair, and verify errors or bottlenecks (e.g., cleaning agents that are associated with wrong repairs) in the cleaning process. Thus, humans are isolated from the specific cleaning logic of a specific cleaning algorithm.

- **Accountability:** There are many factors involved in computing a given repair. Based on human feedback (e.g., human reports an error in a repaired cell), the system should automatically assess the reliability of different factors (e.g., agents, rules, etc.) that were involved in computing a repair over time. This assessment is crucial for humans to identify bottlenecks, i.e., factors associated with inaccurate repairs, in the data cleaning process.

- **Human Cost Optimization:** The system should be able to reason about the expertise of different humans when assigning cleaning tasks. It should also account for the cost and expertise when involving a given human in a cleaning task.

**Organization:** We present the architecture of our envisioned system in Section 2. In Section 3, we discuss key features to characterize humans in the cleaning pipeline. In Section 4, we discuss the problem of automatically assigning humans to cleaning tasks. We present and contrast different cost optimization strategies in Section 5. We address strategies to identify bottlenecks in the data cleaning pipeline in Section 6. We discuss related work in Section 7 and conclude in Section 8

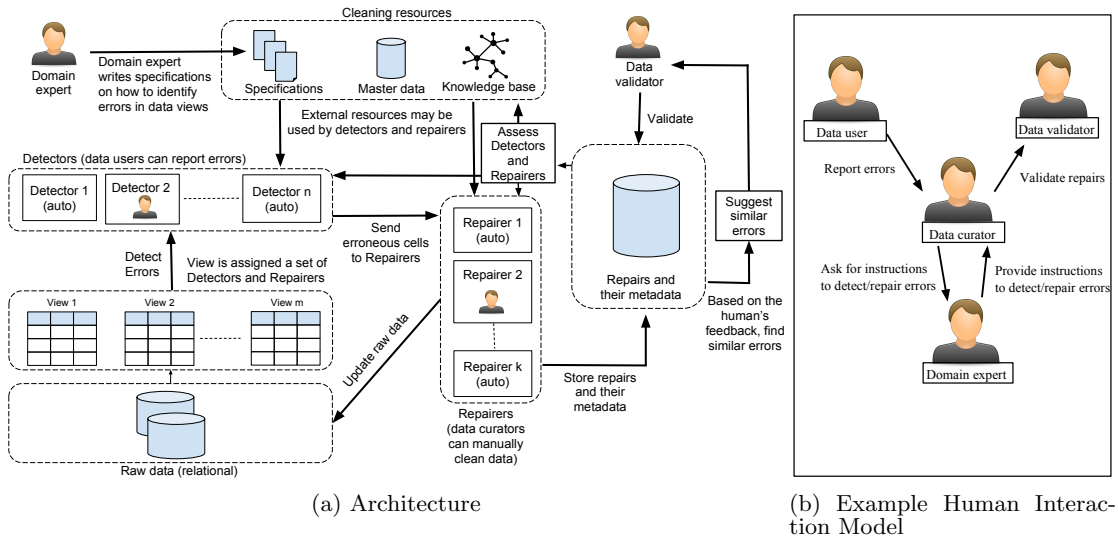

(a) Architecture  (b) Example Human Interaction Model

Figure 2: Architecture (vision) and an example Human Interaction Model

## 2. ARCHITECTURE OVERVIEW

### 2.1 Terminology

Consider a relational database $D$ containing relations $R_1$, $R_2, ..., R_n$. Every relation $R_i$ ($1 \leq i \leq n$) contains a set of attributes $A_1^i, A_2^i, ..., A_k^i$ with domains $dom(A_1^i), dom(A_2^i)$, ..., $dom(A_k^i)$ respectively. For the instance $I_i$ of $R_i$ containing tuples $T$, a cell $c$ is the value of a tuple $t \in T$ in attribute $A \in R_i$, denoted $t[A]$.

**Detector:** Detectors are humans or programs that, given a set of cells as input, provide a set of cells that are potentially erroneous as output. Example detector programs are those that use data quality rules (e.g., Denial Constraints) to identify the cells that violate those rules.

**Repairer:** Repairers are humans or programs that update the input cells in a way that "fixes" the data errors.

**Repair:** We refer to an update to a set of cells $C$ made by a Repairer $R$ as a *repair*.

**Accurate Repair:** A repair is accurate if it contains cells with values that match the ground truth.

### 2.2 Architecture

Figure 2a illustrates a proposed architecture to implement our system vision. In a nutshell, there are four main components: Detectors, Repairers, Cleaning resources and Validators. All the Detectors and Repairers are treated as pluggable black boxes. One could use any number of detection and repairing algorithms to clean the data. Since different agents can be involved to detect/repair different parts of the data, Detectors and Repairers are applied to data views, which contain subsets of the raw data. Furthermore, Detectors and Repairers may use *cleaning resources* such as rules, masterdata, etc., to detect and/or repair the data. Cleaning resources are commonly generated by humans, we explore in Section 6 how the envisioned system should make it easy for humans to identify agents or cleaning resources that produce inaccurate repairs. Finally, in addition to detecting and fixing errors, humans are also able to validate the computed repairs, and based on their feedback, the system assesses the reliability of different factors that were involved in computing the repairs.

**Data Cleaning job:** The envisioned system allows humans to declaratively specify a data cleaning operation as a function of several parameters. Specifically, a data cleaning job is represented as the quadruplet $\langle C, D, R, V \rangle$ where: $C$ is the set of input cells (cannot be empty), $D$ is the set of Detectors to be used to detect errors in $C$, $R$ is the set of Repairers to repair the errors found in $C$, $V$ is the set of humans to validate the produced repairs. Using this representation, we can capture all the cleaning scenarios. For example, if $D$ and $R$ are empty and $V$ is not empty, then, the job will be a validation job of the cells in $C$.

*Example 2.* In Example 1 (scenario 1), the two cleaning jobs are represented as:
$job_1 : \langle C = *, D = \{"Alice"\}, R = \{"Bob"\}, V = \emptyset \rangle$ (C = * indicates that Alice can report an error in any cell).
$job_2 : \langle C = \{tb[Zip] = *, tb[City] = *\}, D = \{\phi_1\}, R = \{R_1\}, V = \emptyset \rangle$

$job_1$ states that "Alice" can report errors in any data cell, and if she does, the error will be repaired by "Bob". The repair that "Bob" performs is not subjected to a validation ($V = \emptyset$).

$job_2$ states that we are using $\phi_1$ (in practice, there is a program that projects $\phi_1$ on the data to extract violations, but for simplicity, we are only including the rule) to detect errors in all the Zip and City cells. The errors are then repaired using $R_1$. Like $job_1$, the repairs are not subjected to a validation ($V = \emptyset$).

## 3. HUMANS IN THE CLEANING PROCESS

While several research efforts involve the human in specific cleaning problems (e,g,. Entity Resolution [20], Integrity Constraints [22], Data Fusion [15]), there is no proposal that involves humans for general data cleaning (regardless of the cleaning problem at hand). Furthermore, characterizing human expertise for the purpose of general data cleaning remains unexplored. Particularly, data cleaning efforts that use crowdsourcing [20, 6] assume that crowd workers are non-experts. On the other end of the spectrum, we have data cleaning methods [22, 18] that assume humans are experts whose feedback is assumed to be always correct. In practice, humans can have different degrees of expertise on different parts of the data. We shed some light to highlight key challenges that need to be addressed to realize this characterization.

### 3.1 Characterizing Human Expertise

**Cleaning tasks:** Humans interact in various ways in the cleaning process. Based on our vision, we list four human-driven tasks, referred to in this paper by *cleaning tasks*, that a human-centric data cleaning system needs to support.

1. **Detection:** Humans should be able to report errors in a given set of cells.

2. **Repairing:** When errors are reported, humans should be able to fix those errors by updating the data to reflect accurate values.

3. **Validation:** Humans should be able to verify a repair that has been made by another cleaning agent (human or automatic).

4. **Specification:** Humans should be allowed to write specifications (e.g., FD rules) to detect data errors.

**Human roles:** The above interaction cases impose a distinction between different human roles in the cleaning process. Human roles in data cleaning are not well-studied. At a high level, we can think of a separation between human roles based on the knowledge of the data, the domain, and the technical tools needed to update and transform the data. In the Detection task, the person reporting errors does not have to be technical. We refer to this person as the *Data User*. In the Repair task, because the human has to update the data with new values, the human has to possess the necessary technical background to perform the repair without introducing new data errors for technical reasons (e.g., a faulty SQL statement), we refer to this person as the *Data Curator*. In order to know *how* to repair a reported error, the *Data Curator* has to be knowledgeable enough about the erroneous data. In the Validation task, the humans validating the repairs have to be knowledgeable about the data they are asked to validate but do not have to be technical. For example, the system can ask a person yes/no questions about some data cells. We refer to this person as the *Data Validator*. In the Specification task, the person has to be a domain expert who can write specifications (for example, in the form of rules) that are then used to capture data errors. We refer to this person as the *Domain Expert*.

**Data Expertise:** Humans have different knowledge about different parts of the data. For example, a person working in the Sales department is probably more aware of the Sales data than someone working in the Marketing department. When assigning humans to cleaning tasks, it is important the system makes sure the assigned humans are knowledgeable enough in the data they are asked about. In a human-centric data cleaning system, every human has a history of the data cells they helped clean. Through the *Validation* task, the system can learn how *good* a given human is for a certain cleaning task and for a given cell. For example, a simple measure to quantify the expertise of a human $h$ on data cells $C$ is the following:

$$Expertise(h, C, T) = \frac{\#correct(C, T)}{\#validated(C, T)} \quad (1)$$

Equation 1 calculates the ratio of correct cells in a given task for the cells $C$ ($correct(C, T)$) over to the number of cells in $C$ that were subject to validation.

For example, we would like to measure the expertise of a human $h$ in the detection task for a set of cells $C$. Assume $h$ correctly reported errors in two of these cells. The total number of cells that were validated by a Data Validator in $C$ is 4. Therefore: $Expertise(h, C, Detect) = \frac{2}{4} = 0.5$

**Cost Model.** Involving humans is generally expensive. It is important to be able to characterize the cost of involving a human to perform a certain cleaning task. For example, involving domain experts is generally more costly than involving ordinary data users. This cost should also take into consideration the availability of different human roles. For instance, if we have very few data curators, we would want to make sure they are assigned the most critical tasks only. Furthermore, it would be interesting to incorporate the cognitive effort of looking at the data to perform a cleaning task.

**Human Budget.** The human budget for a data cleaning job $j$ could be expressed as a combination of many factors including the maximum number of humans available to perform a certain task, the total money cost to spend to perform a task, etc.

We are now ready to formally define a Human characterization of a human $h$.

*Definition 1.* Human Characterization. A human $h$ in a data cleaning scenario is represented as h: ⟨$Role$, $Data$, $Cost$, $Expertise$⟩ where $Role$ is the role of the human, $Data$ is the set of cells $h$ is knowledgeable about, $Cost$ is the cost of involving h, $Expertise$ is a score that reflects how good $h$ is for the role $Role$ in cells $Data$.

## 4. TASK ALLOCATION

The envisioned system should allow users to define data cleaning jobs without explicitly stating the humans involved. Specifically, the system should be able to select from a pool of humans $H$ with different characterizations, the right human for the right task. An example jobs is defined as follows:
$job_3 = \langle C = *, D = \{\text{``Alice''}\}, R = H, V = \emptyset \rangle$

In $job_3$, The set of repairers is set to the set of humans $H$ available. This makes the system responsible for assigning a repairer for the errors that *Alice* detects. In our system, we are only interested in automatically assigning humans to cleaning tasks. Assigning automatic agents to cleaning tasks is outside the scope of the proposed vision.

Given a set of humans with their characterizations, the proposed system should be able to automatically assign cleaning tasks to them. We now discuss key building blocks that are needed to effectively assign cleaning tasks to humans.

### 4.1 Interaction Between Humans

It is crucial to develop an interaction model between different human roles to optimize the cleaning effort. Ideally,

we should aim for an interaction model that produces the best cleaning results at the least human cost. In particular, a "good" interaction model should: (1) Minimize the communication overhead between different human roles; and (2) Account for all possible human-to-human interaction cases in the cleaning scenario. These cases are dictated by the set of human roles and their expertise. For instance, as illustrated in Figure 2b and using the roles we defined previously, the possible human-to-human interaction scenarios are the following:

- Data User reports errors to the Data Curator.
- Domain Expert provides specifications (e.g., rules, etc.) to the Data Curator to enforce on the data.
- Data Curator reports errors found in specifications to the Domain Expert.
- Data Validator validates fixes performed by the Data Curator.

### 4.2 Task Assignment

Given a data cleaning job $j$ for cells $C$, a pool of humans, say $H$, and a budget, say $B$, the framework should assign automatically cells in $C$ to humans in $H$ (e.g., $job_3$ defined above). The assignment should guarantee the following properties: (1) **Coverage:** If the job is to be performed by humans only, every cell in $c$ should be covered by at least one human; (2) **Maximize expertise**: The assigned humans to given cells should have good knowledge about these cells; (3) **Minimize cost:** The human cost should not exceed Budget $B$.

*Example 3.* Assume that we have a validation task on all the $Sal$ cells of Table $Employees\_R$ in Figure 1 defined as the data cleaning job $job_4$ as follows:

$job_4 = \langle C = Employees\_R[Sal], D = \emptyset, R = \emptyset, V = H \rangle$

Consider a pool of humans $H = \{Alice, Bob, \text{and } Sam\}$, and a human budget ($B = 1$) for $job_4$ expressed (for simplicity) as the maximum number of humans involved in the task. $Alice$, $Bob$, and $Sam$ have good knowledge on the following sets of cells $\{te'_1[Sal], te'_2[Sal], te'_3[Sal], te'_4[Sal], te'_5[Sal]\}$, $\{te'_3[Sal], te'_4[Sal]\}$, and $\{te'_5[Sal]\}$, respectively. In this scenario, the system should assign $job_4$ to $Alice$ only (since $B = 1$ and $Alice$ covers all the cells of $Sal$).

## 5. CROSS-AGENT COST OPTIMIZATION

Minimizing the human cost to repair the data has been the cornerstone of numerous research efforts [22]. However, when the human is not aware of the cleaning algorithm's logic, it becomes hard to achieve this goal. For instance, consider a Detector $Dedup$ that detects duplicate records using a clustering algorithm. Figure 3 illustrates a set of data points projected into a 2-dimensional space. $Dedup$ uses some similarity measure $Sim$ to decide if a set of data points belong to the same cluster. Figure 3 shows two clusters $P$ and $Q$ (represented as dashed circles), data points that are in the same cluster are duplicates. Knowing how $Dedup$ works, if we want to validate the output of $Dedup$, we would ask the human to verify if $p_1$ and $q_1$ are not duplicates since those two points are the closest two points (using $Sim$) between $P$ and $Q$. If $p_1$ and $q_1$ are verified to be distinct, then the points in $Q$ are distinct from those in $P$. Without knowing the logic of $Dedup$, we cannot come up with such optimizations of human efforts. In the case of our system, $Dedup$ would simply provide its output as $P$ and $Q$ expressed in terms of records (duplicate records would share the same value of a designated attribute). In this case, involving the human usefully becomes more challenging.

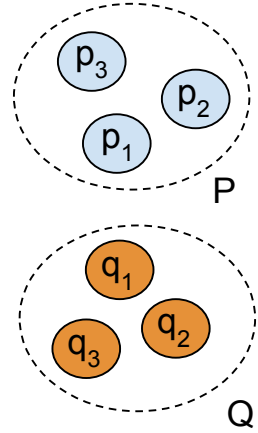

Figure 3: Example clusters $Q$ and $P$ produced by a duplicate detector

We need to answer the following the questions: (1) When we have human and automatic cleaning agents, what are the consequences of involving one over the other on human cost and data quality? (2) Given multiple humans that are assigned the same set of cells to repair, which human do we choose? (3) How do we schedule different cleaning jobs in order to achieve an optimal human cost and data quality?

### 5.1 Quantitative Cost Optimization

When we have humans and automatic agents that are assigned overlapping input data, which one should we prioritize? and what are the consequences for each choice? For example, in Example 1, what if $Bob$ has also been assigned to repair cells $tb_2[City]$, $tb_3[City]$, $tb_2[Zip]$, $tb_3[Zip]$. In this case, the input to $R_1$ (automatic agent) overlaps with the input to $Bob$. This overlap is possible in practice. For example, one may want an automatic agent to clean a large amount of data while requiring the human to repair only a small subset of it. If we want to minimize human intervention, we can simply prioritize automatic agents over humans for a given set of cells. As a result, because $Bob$ is assigned cells that are part of the input to $R_1$, we can simply save cost by not asking $Bob$ on $tb_2[Zip]$, $tb_2[City]$, $tb_3[Zip]$, $tb_3[Zip]$, but we would still ask him to fix the Salary value because Salary is not input to the automatic agent. While human intervention is minimized in this strategy, we note the following points:

- Human cost is minimized at the expense of data quality. That is, humans generally perform better repairs than automatic agents.
- This strategy can be suitable if the automatic agents provide high repair accuracy.

### 5.2 Qualitative Cost Optimization

This strategy gives preference to humans over automatic tools. As a result, the human cost will be higher compared to the previous strategy. In this strategy, when the input cells for an automatic agent overlap with those for a human agent, the system first invokes the automatic agent, and then asks the human to correct the overlapping cells. This way, the human updates will be ordered last and will not be undone by the automatic agent. Using this strategy, we note the following:

- Because humans are prioritized over automatic agents, it is expected that this strategy results in better data quality compared to the previous one.

- Human cost is high in this strategy. This strategy is suitable when invoking the automatic agents would result in low accuracy in terms of repair quality.

## 6. IDENTIFICATION OF BOTTLENECKS

There are several factors involved in repairing a cell, say $c$. We refer to these factors by $factors(c)$, where they include: (1) Detectors: Human or automatic agents that have flagged $c$'s old value as erroneous; (2) Repairers: Agents (humans or automatic) that have computed the repair in $c$; (3) Cleaning resources: Resources used to compute $c$, where they may include Rules, Metadata, etc. (4) Data Validators: Humans that validated $c$ as a correct repair (if $c$ has been subject to validation).

After human validation, if a repair for $c$ is deemed accurate (respectively, inaccurate), then every factor in $factor(c)$ should be rewarded (respectively, penalized). Providing this accountability will help identify factors that are commonly associated with inaccurate repairs. This assessment is crucial for humans as it helps them identify bottlenecks in the cleaning pipeline as a whole.

**Scoring factors:** One way to capture the quality of different factors is to compute a score for each one of them that reflects how "good" they are. A simple way to capture the quality of a given factor $f \in factor(c)$ is the following:

$$Quality(f) = \frac{\#correct(f)}{\#validated(f)} \quad (2)$$

Equation 2 calculates the ratio of correct cells (as validated by a human) where $f$ was involved over the total number of validated cells where $f$ was involved.

Since they would result in inaccurate repairs, low-quality factors would elicit more human feedback than high-quality ones. Therefore, identifying them is crucial to minimize the human cost in the cleaning process.

<u>Scenario 2:</u> Consider Example 1. We want to perform a new cleaning iteration with an additional FD rule that will be enforced on table $Branches\_R$ (Figure 1). Let us add an *incorrect* FD rule: $\phi_2 : City \rightarrow Zip$. This rule states that records that share the same City should have the same Zip code. This is in reality not correct because a city can have multiple zip codes. We now create a new data cleaning job: $job_5 : \langle C = \{tb'[Zip] = *, tb'[City] = *\}, D = \{\phi_1, \phi_2\}, R = \{R_1\}, V = \emptyset \rangle$

The set of violating cells will be $C^{\not\models} = \{tb'_1[Zip], tb'_1[City], tb'_4[Zip], tb'_4[City], tb'_5[Zip], tb'_5[City]\}$. Let us assume that $R_1$ lifts the violation by setting $tb'_1[Zip] = 47904$. Let us call the repaired cell $tb''_1[Zip]$.

If we want to ask $Jen$, a human data validator about the repairs computed by $R_1$, which violating cells should we ask her to validate their repair? More importantly, how does the choice of cells we choose to validate affect our ability to isolate troublesome factors? Furthermore, how can we adjust the choice of cells to validate to our available human budget? To shed some light on answering those questions, we discuss the following key cases:

1. If we validate cells that were computed using many factors, we get an *aggregate* feedback on all the involved factors. For instance, asking $Jen$ to validate $tb'_5[City]$ would provide a feedback about $\phi_1$, $\phi_2$ and $R_1$ (that cell was involved in two violations across different cleaning iterations). This is useful to get a feedback about many factors at once, however, it may not be good at isolating factors to identify the bottlenecks causing inaccurate repairs. This strategy is suitable when the cost of involving human validators is high. Therefore, this strategy allows us to have an idea about as many factors as possible using the least number of cells to hopefully identify a combination of factors that produced inaccurate repairs.

2. If we validate cells that were computed using few factors, we get a more fine-grained feedback about the involved factors. For example, asking $Jen$ to validate $tb''_1[City]$ and $tb''_1[Zip]$ (which represent the repairs for cells $tb'_1[City]$ and $tb'_1[Zip]$ respectively) would isolate $\phi_2$ as a problematic FD rule (since $tb'_1[Zip]$ and $tb'_1[City]$ violated $\phi_2$ only). While this strategy provides a better isolation of factors, it involves more cells to be validated which translates into spending a higher human cost.

## 7. RELATED WORK

There is a rich literature on Data Cleaning techniques and theory [1, 12, 10]. We discuss a few papers in two areas: general data cleaning systems and human-assisted data cleaning techniques.

**General Data Cleaning Systems:** A strongly related system to our proposal is the data cleaning system NADEEF [8]. Like our envisioned system, NADEEF adopts a system-approach to realize an end-to-end data cleaning framework that supports a number of data cleaning problems (Integrity Constraints, Deduplication, etc.). NADEEF offers a programming interface so that users can implement detection and repairing components. As opposed to NADEEF, our framework not only supports automatic agents, but also involves humans in the data cleaning pipeline. Another related system is $KATARA$ [6] which leverages the crowd and knowledge bases (KB) to clean dirty tables. $KATARA$ is not rule-driven and can repair any cells in the input tuples (hence its categorization as a general data cleaning system). $KATARA$ jointly uses the KB and the crowd feedback to identify correct and erroneous data in the input dirty table. Our vision is different from $KATARA$ as we are considering humans with different expertise and roles. Furthermore, our envisioned system supports any number of cleaning agents for different cleaning problems (integrity constraints, deduplication, etc.).

**Human-assisted Data Cleaning:** The idea of assisting humans in specific data cleaning problems (Entity Resolution, Schema transformations, etc.) is not new [16, 22, 19, 18, 6, 3]. However, human involvement is usually coupled to the underlying cleaning logic. This is different from our framework where cleaning algorithms are treated as pluggable black boxes and the human only interacts with their output.

The crowdsourcing literature [14] focuses on involving the crowd, which is typically assumed to contain non-expert humans to solve specific problems [19, 20, 11, 6]. This body of work is important to study to realize the full potential of the proposed framework. However, in our framework, humans have multiple roles with varying degrees of expertise. Furthermore, our setup supports non-human agents as well.

## 8. CONCLUSIONS

Scaling the generation and processing of big data often comes at the cost of its quality. We presented our vision for a framework that assists humans in all the major stages of the cleaning pipeline. We proposed several properties that need to be met to unlock the full potential of a given data cleaning scenario. We raised several questions that need to be addressed in order to bring such a vision to life. There are still many other questions that were not addressed in this vision such as data privacy, i.e., how can humans clean the data in the presence of privacy constraints? But we believe the proposed vision raises the central questions that we need to answer first to realize a human-centric data cleaning system.